\documentstyle[preprint,prl,aps]{revtex}
\tightenlines
\begin{document}
\preprint{
\vbox{\halign{&##\hfil\cr
        & OHSTPY-TH-98-001   \cr
        & hep-ph/9801226     \cr
        & January 6, 1998  \cr}}}

\title{An Explanation for the $\rho \,$--$\pi$ Puzzle \\
        of $J/\psi$ and $\psi'$ Decays}

\author{Yu-Qi Chen and Eric Braaten}
\address{Physics Department, Ohio State University, Columbus OH 43210}

\maketitle
\begin{abstract}
We propose a new explanation for the longstanding puzzle
of the tiny branching fraction of  
$\psi' \to \rho \pi$ relative to that for $J/\psi \to \rho \pi$.
In the case of $J/\psi$, we argue that this
decay is dominated by a higher Fock state in which the $c \bar c$ pair
is in a color-octet  $^3S_1$ state
and via annihilate process $c \bar c \to q \bar q$.
In the case of the $\psi'$, we argue that the probability
for the $c \bar c$ pair in this higher Fock state to be close enough 
to annihilate is suppressed by a dynamical effect related to the 
small energy gap between the mass of the $\psi'$ and the 
$D \bar D$ threshold.
\end{abstract}

\vfill \eject

\narrowtext

A long-standing mystery of charmonium physics is the 
``$\rho \,$--$\pi$ puzzle''
of $J /\psi$ and $\psi'$ decays.  These particles are nonrelativistic 
bound states of a charm quark and its antiquark.
Their decays into light hadrons are
believed to be dominated by the annihilation of the $c \bar c$ pair into 
three gluons.  In order to annihilate, the $c$ and $\bar c$ must have a 
separation of order $1/m_c$, which is much smaller than the size of the 
charmonium state.  Thus the annihilation amplitude 
for an S-wave state like $J/\psi$ or $\psi'$ must be proportional 
to the wavefunction at the origin, $\psi({\bf r}=0)$.  
The width for decay into any specific final state $h$ 
consisting of light hadrons is therefore proportional to $|\psi(0)|^2$.  
The width for decay into $e^+e^-$ is also proportional to $|\psi(0)|^2$.
This leads to the simple prediction that the ratio of the 
branching fractions for $\psi'$ and $J/\psi$ is given by the
``$15\%$ rule'':
\begin{equation}
Q_h \;\equiv\; 
{B (\psi' \to h) \over B(J / \psi \to h)} \;=\; 
Q_{ee} \;=\; 
(14.7 \pm 2.3) \% \,. 
\label{Q-h}
\end{equation}
The $\rho \,$--$\pi$ puzzle is that the prediction (1) 
is severely violated in the $\rho \pi$ and several other decay channels.  
The first evidence for this effect
was presented by the Mark II collaboration in 1983 \cite{MarkII}.  
They found that $Q_{\rho \pi} < 0.6\%$ and $Q_{K^*K} < 2 \%$.  
Recent data from the BES collaboration has made the puzzle even sharper.  
They obtained $Q_{\rho \pi} < 0.23\%$, $Q_{K^{*+}K^-} < 0.64 \%$, and
$Q_{K^{*0} \bar K^0} = (1.7 \pm 0.6)\%$ \cite{BES}.  
Thus the suppression of $Q$ relative to
the $15\%$ rule is about 10 for ${K^*}^0 \bar K^0$ and greater than 65 for 
$\rho \pi$.

A summary of the proposed explanations of the $\rho \,$--$\pi$ puzzle has 
recently been given by Chao \cite{Chao}.  
Hou and Soni \cite{H-S} suggested that
$J/\psi \to \rho \pi$ is enhanced by a mixing of the $J/\psi$ with a 
glueball ${\cal O}$ that decays to $\rho \pi$.
Brodsky, Lepage, and Tuan \cite{B-L-T} emphasized that
$J/ \psi \to \rho \pi$ violates the helicity selection rule of 
perturbative QCD, and argued that the data requires
${\cal O}$ to be narrow and nearly degenerate with the $J/\psi$.  
Present data from BES constrains the mass and width of the glueball 
to the ranges $|m_{\cal O} - m_{J/\psi}| < 80$ MeV and 4
MeV $< \Gamma_{\cal O} < $ 50 MeV \cite{Hou}.
This mass is about 700 MeV lower than the lightest 
$J^{PC} = 1^{--}$ glueball observed in lattice simulations of QCD 
without dynamical quarks \cite{Peardon}.
There are explanations of the $\rho \,$--$\pi$ puzzle that 
involve the dependence of the decay amplitude on the energy 
of the charmonium state.  
Karl and Roberts \cite{K-R} suggested that the decay proceeds
through  $c \bar c \to q \bar q$ followed by the fragmentation of the 
$q \bar q$ into $\rho \pi$.  They argued that the fragmentation 
probability is an oscillatory function of the energy which could have a
minimum near the mass of the $\psi'$.  
Chaichian and Tornqvist \cite{C-T} pointed out that the suppression of 
$\psi'$ decays could be explained if the form factors 
for two-body decays fall exponentially with the energy
as in the nonrelativistic quark model.
There are other explanations of the $\rho \,$--$\pi$ puzzle that 
rely on the fact that there is a node in the radial wavefunction 
for $\psi'$,  but not for $J/\psi$.  
Pinsky \cite{Pinsky} suggested that this node makes 
$\psi' \to \rho \pi$ a ``hindered M1 transition'' like 
$J/\psi \to \eta_c \gamma$.  
Brodsky and Karliner \cite{B-K} suggested that the decay
into $\rho \pi$ proceeds through intrinsic charm components 
of the $\rho$ and $\pi$ wavefunctions.  They argued that the $c \bar c$ pair 
in the $|u \bar d c \bar c \rangle$ Fock state of the $\rho^+$ or $\pi^+$ 
has a nodeless radial wavefunction which gives
it a larger overlap with $J/\psi$ than $\psi'$.
Finally, Li, Bugg, and Zou \cite{L-B-Z} have
suggested that final-state interactions involving the rescattering of 
$a_1 \rho$ and $a_2 \rho$ into $\rho \pi$ could be important
and might interfere destructively in the case of the $\psi'$.  

In this Letter, we present a new explanation of the $\rho \,$--$\pi$ puzzle.  
We argue that the decay $J/\psi \to \rho \pi$ is 
dominated by a higher Fock state of the $J/\psi$ in which the $c \bar c$ 
is in a color-octet $^3S_1$ state.  The $c \bar c$ pair in this Fock state 
can annihilate via $c \bar c \to g^* \to q \bar q$.
The amplitude for forming $\rho \pi$ is dominated by the endpoint of the 
meson wavefunctions, where a single $q$ or $\bar q$ carries most of the 
momentum of the meson.
The suppression of $\psi' \to \rho \pi$ is attributed to a suppression 
of the $c \bar c$ wavefunction at the origin for the  
higher Fock state of the $\psi'$.  
Such a suppression can arise from a dynamical effect associated with
the small energy gap between the mass of the $\psi'$ and 
the $D \bar D$ threshold.

It is convenient to analyze charmonium decays
using nonrelativistic QCD (NRQCD) \cite{B-B-L}, 
the effective field theory obtained 
by integrating out the energy scale of the charm quark mass $m_c$.
In Coulomb gauge, a charmonium state has a Fock state decomposition 
in which the probability of each Fock state scales as a definite power of $v$, 
the typical relative velocity of the charm quark.
The dominant Fock state of the $J/\psi$ and the $\psi'$ is
$| c \bar c_1(^3S_1) \rangle$, whose probability ${\cal P}$ is of order 1.
We denote the color state of the $c \bar c$ pair by a subscript
(1 for color-singlet, 8 for color-octet) and we put the angular momentum 
quantum numbers in parentheses.  The Fock states
$| c \bar c_8(^3P_J) +  S \rangle$,
where $S$ represents dynamical gluons or 
light quark-antiquark pairs with energies of order $m_c v$ or less,
have probability ${\cal P} \sim v^2$.
The next most important Fock states include
$| c \bar c_8(^1S_0) +  S \rangle$ with ${\cal P} \sim v^3$
(or ${\cal P} \sim v^4$ if perturbation theory is sufficiently accurate
at the scale $m_c v$) and
$| c \bar c_8(^3S_1) +  S \rangle$ with ${\cal P} \sim v^4$.

The decay of the $J/\psi$ into light hadrons proceeds via
the annihilation of the $c$ and $\bar c$, which  
can occur through any of the Fock states.  
This is expressed in the NRQCD factorization formula 
for the inclusive decay rate \cite{B-B-L}, 
which includes the terms
\begin{eqnarray}
\Gamma (J/\psi \to {\it l.h.} ) 
&=& 
\left( {20 (\pi^2-9) \alpha_s^3 \over 243 m_c^2} 
	+ {16 \pi \alpha^2 \over 27 m_c^2} \right)
	\langle {\cal O}_1(^3S_1) \rangle_{J/\psi}
\;+\; {5 \pi \alpha_s^2 \over 6 m_c^2} 
	\langle {\cal O}_8(^1S_0) \rangle_{J/\psi}
\nonumber \\
&&
\;+\; {19 \pi \alpha_s^2 \over 6 m_c^2} 
	\langle {\cal O}_8(^3P_0) \rangle_{J/\psi} 
\;+\; {\pi \alpha_s^2 \over m_c^2} 
	\langle {\cal O}_8(^3S_1) \rangle_{J/\psi}
	\;+\; \ldots \,. 
\label{fact}
\end{eqnarray}
where $\alpha_s = \alpha_s(m_c)$.
The matrix elements are expectation values in the $J/\psi$ of local 
gauge-invariant NRQCD operators that measure the inclusive probability 
of finding a $c \bar c$ in the $J/\psi$ at the same point and in 
the color and angular-momentum state specified.
The matrix element of the $c \bar c_1(^3S_1)$ term in (\ref{fact}) is 
proportional to the square of the wavefunction at the origin and
scales as $v^3$.  Its coefficient includes a term of order 
$\alpha_s^3$ from $c \bar c \to ggg$ and a term of order $\alpha^2$ 
from the electromagnetic annihilation process 
$c \bar c \to \gamma^* \to q \bar q$.
The color-octet terms in  (\ref{fact}) represent contributions 
from higher Fock states.  Their matrix elements scale 
like $v^6$, $v^7$ and $v^7$, respectively.
Their coefficients are all of order $\alpha_s^2$
and come from $c \bar c \to g g$ for 
the $c \bar c_8(^1S_0)$ and $c \bar c_8(^3P_0)$ terms  
and from $c \bar c \to g^* \to q \bar q$ for the $c \bar c_8(^3S_1)$ term.
Note that the coefficients of the color-octet matrix elements
are two orders of magnitude larger than that of
$\langle {\cal O}_1(^3S_1) \rangle_{J/\psi}$, which suggests
that the higher Fock states may play a more important role in 
annihilation decays than is commonly believed.
 
Bolz, Kroll, and Schuler have emphasized that contributions from higher 
Fock states should also be important in exclusive decays of charmonium 
into light hadrons \cite{B-K-S}.  
They argued that the amplitude for a two-body annihilation decay 
such as $J/\psi \to \rho \pi$ satisfies a factorization formula 
which separates the scale $m_c$ from the lower
momentum scales.
The decay amplitude is expressed 
in terms of hard-scattering factors $\widehat {\cal T}$ 
that involve only the scale $m_c$,
initial-state factors $\cal I$ that involve scales of order $m_c v$ and
lower, and final-state factors $\cal F$ 
that involve only the scale $\Lambda_{\rm QCD}$. 
If $m_c$ was asymptotically large, the dominant terms in the factorization 
formula would have the minimal number of partons involved in the 
hard scattering.  Terms involving additional soft partons in the 
initial state are suppressed by powers of $v$.  Terms involving 
additional hard partons in the final state are suppressed by powers of 
$\Lambda_{\rm QCD}/m_c$.  Thus the asymptotic factorization formula 
has the schematic form
\begin{equation}
{\cal T}(J/\psi \to \rho^+ \pi^-) 
= \sum_{u \bar d, d \bar u} 
\left[ \widehat {\cal T} (c \bar c_1(^3S_1) \to u \bar d + d \bar u)
	{\cal F}_{\rho^+}(u \bar d) \; {\cal F}_{\pi^-}(d \bar u) \right]
{\cal I}_{J/\psi} (c \bar c_1(^3S_1)).
\label{fact-asymp}
\end{equation}
The initial-state factor is proportional to the wavefunction at the origin
for the $| c \bar c_1(^3S_1) \rangle$ state.
The lowest order contributions to the hard-scattering factor are of order 
$\alpha_s^3$ from the annihilation of $c \bar c$ into three virtual gluons
and $\alpha \alpha_s$ from the annihilation into a virtual photon.
As pointed out by Brodsky and Lepage \cite{Brodsky-Lepage},
the leading term in the asymptotic factorization formula (\ref{fact-asymp})
is strongly constrained by the vector character of the QCD interaction
between quarks and gluons.  It vanishes unless the sum of the helicities 
of the mesons is zero.  Rotational symmetry then requires
the angular distribution to be $1 - \cos^2 \theta$.
This helicity selection rule is violated by the decay into $\rho \pi$.
Parity and rotational symmetry require the helicity of the $\rho$ 
to be $\pm 1$ and the angular distribution to be $1 + \cos^2 \theta$.
Since the helicity of the pion is 0, 
the helicity selection rule is violated.
Thus the amplitude for $J/\psi \to \rho \pi$
is suppressed by a factor of 
$\Lambda_{\rm QCD}/m_c$ relative to that for some other mesons.

Since the charm quark mass $m_c$ is less than an order of magnitude larger
than $\Lambda_{\rm QCD}$, there can be large corrections to the 
asymptotic decay amplitude (\ref{fact-asymp}).  
In particular, there can be significant regions 
of phase space in which some of the gluons involved in the hard
scattering are relatively soft.  It might therefore be more appropriate
to absorb them into the initial-state or final-state factors.
In this case, not all of the soft partons in $S$ need be involved in the 
hard scattering, and not all the partons that form the $\rho^+$ and $\pi^-$
need be produced by the hard scattering.  For example, 
there can be a contribution from the Fock state
$| c \bar c_8(^3S_1) + S \rangle$  that involves the hard-scattering 
process $c \bar c \to u \bar u$.  This produces a state 
$| u \bar u + S \rangle$ that consists of the soft partons $S$ 
together with a $u$ and a $\bar u$ that are back-to-back and whose
momenta are approximately $m_c$.  Such a state has a nonzero overlap
with the final state $| \rho^+ \pi^- \rangle$. 
The overlap comes from the endpoints of the meson wavefunctions, 
in which most of the momentum of the $\rho^+$ is carried by the 
$u$ and most of the momentum of the $\pi^-$ is carried by the $\bar u$.
There is also an endpoint contribution involving the dominant 
$| c \bar c_1(^3S_1) \rangle$ Fock state, which can annihilate
into $q \bar q$ through a virtual photon.
These contributions to the $T$-matrix element can be written 
schematically in the form
\begin{eqnarray}
{\cal T}(J/\psi \to \rho^+ \pi^-)
&=& \sum_{q \bar q} \;
\widehat{\cal T} (c \bar c_8(^3S_1) \to q \bar q) \;
\sum_S \left[ {\cal F}_{\rho^+ \pi^-}(q \bar q + S) \;
	{\cal I}_{J/\psi} (c \bar c_8(^3S_1) + S) \right] 
\nonumber \\
&& \;+\; \sum_{q \bar q} \; 
\widehat {\cal T} (c \bar c_1(^3S_1) \to q \bar q) \;
	{\cal F}_{\rho^+ \pi^-}(q \bar q) \;
	{\cal I}_{J/\psi} (c \bar c_1(^3S_1)) \;,
\label{fact-endpt}
\end{eqnarray}
where the sum over $q \bar q$ includes $u \bar u$ and $d \bar d$.
In the factorization formula (\ref{fact}) for inclusive decays,
the two contributions on the right side of  (\ref{fact-endpt}) 
contribute to the $c \bar c_8(^3S_1)$
and $c \bar c_1(^3S_1)$ terms, respectively.

The endpoint contribution in (\ref{fact-endpt}) leads to a definite 
angular distribution.  Since the $q$ and $\bar q$ carry most of the 
momenta of the mesons,  the angular distribution
of the mesons will follow
that of the $q$ and $\bar q$, which is $1 + \cos^2 \theta$.  
Thus (\ref{fact-endpt}) will contribute most strongly to form factors which 
allow the angular distribution $1 + \cos^2 \theta$.
It will also contribute most strongly to decays into mesons
like $\rho$ and $\pi$ for which most of the momentum can be carried by a 
single $q$ or $\bar q$.
There are also endpoint contributions involving the 
hard-scattering process $c \bar c_8(^3P_J) \to g g$,
which produces a pair of hard gluons with the angular distribution
$2 + \cos^2 \theta$, and $c \bar c_8(^1S_0) \to g g$, 
for which the angular distribution is isotropic.
Their contributions to $J/\psi \to \rho \pi$ are suppressed
by the small probabilities for most of the momentum of the 
$\rho$ or $\pi$ to be carried by a single gluon
and by the mismatch between the angular distribution of the gluons and 
the $1 + \cos^2 \theta$ distribution of the $\rho \pi$.

We argue that the color-octet term in (\ref{fact-endpt})
may actually dominate the decay rate for $J/\psi \to \rho \pi$. 
We compare the various factors in that term with those in
the asymptotic expression (\ref{fact-asymp}).  
The hard-scattering factor $\widehat{\cal T}$ in the color-octet 
term in (\ref{fact-endpt}) is only of order $\alpha_s$,
compared to $\alpha_s^3$ in (\ref{fact-asymp}).  
The suppression from the initial-state factor in (\ref{fact-endpt}), 
including the sum over $S$, might be as little as a factor of $v^2$
relative to that in (\ref{fact-asymp}) if the additional soft partons $S$
consist of a single gluon with momentum of order $m_c v^2$.
As for the final-state factors, (\ref{fact-endpt}) is suppressed
by the endpoints of the meson wavefunctions, while
(\ref{fact-asymp}) is suppressed by $\Lambda_{\rm QCD}/m_c$ from the 
violation of the helicity selection rule.
Considering the various suppression factors, it is certainly 
plausible that the $c \bar c_8(^3S_1)$ term
in (\ref{fact-endpt}) could dominate 
over the leading contribution from the $c \bar c_1(^3S_1)$ Fock state.  

We have argued that the decay $J/\psi \to \rho \pi$ may be dominated by the 
annihilation of the $c \bar c$ pair in the $| c \bar c_8(^3S_1) + S \rangle$ 
Fock state via $ c \bar c \to q \bar q$.  If this is true,
then the $\rho \,$--$\pi$ puzzle can be explained by a suppression of this 
decay mechanism in the case of the $\psi'$.  
This suppression can arise from the 
initial-state factor ${\cal I}_{J/\psi} (c \bar c_8(^3S_1) + S)$
if the wavefunction for the 
$| c \bar c_8(^3S_1) + S \rangle$  Fock state is suppressed in the region
in which the separation of the $c \bar c$ is less than or of order 
$1/m_c$.    Note that it does not require a suppression of the probability 
for the higher Fock state, but just a shift in the probability
away from the region in which the $c$ and $\bar c$ are close enough to
annihilate.  A possible mechanism for this suppression 
is a dynamical effect related to the small energy gap between 
the mass of the $\psi'$ and the $D \bar D$ threshold.  
In the Born-Oppenheimer approximation\cite{J-K-M},   
the $c\bar{c}$ pair in the dominant $|c\bar{c}_1 \rangle $
Fock state  moves adiabatically in response to a potential $V_1(R)$  
given by  the minimal energy of  QCD in the 
presence of a color-singlet $c\bar{c}$ pair with fixed separation $R$.
Similarly,  the $c \bar c$ pair in the 
$| c \bar c_8(^3S_1) + S \rangle$ Fock state
moves adiabatically in response to a potential $V_8(R)$ 
given by the minimal energy of the soft modes $S$ 
in the presence of a color-octet $c \bar c$ pair with fixed separation $R$. 
At short distances, this potential approaches a repulsive 
Coulomb potential $\alpha_s/(6R)$. At long distances, 
the minimal energy state consists  of $D$ and $\bar{D}$ 
mesons separated by a distance $R$, 
and $V_8(R)$ therefore approches a constant $2(M_D - m_c)$ equal to the
energy of the $D \bar D$ threshold. 
A charmonium state spends most of the time on the color-singlet adiabatic
surface, but it occasionally makes a transition to the color-octet adiabatic
surface. Since the $J/\psi$ is 640 MeV below the $D \bar D$ threshold,
a $c\bar{c}$ pair on the color-octet adiabatic surface is far off  
the energy shell. 
The time spent by the $J/\psi$ on this surface  is therefore too short 
for the $c \bar c$ pair to respond to the repulsive short-distance potential.
The shape of the $c \bar c$ wavefunction for the 
$|c \bar c_8 +S \rangle$ Fock state should therefore be similar 
to that of the $|c \bar c_1 \rangle$ Fock state, which peaks at the origin.
However the mass of the $\psi'$ is only 43 MeV below $D^+ D^-$
threshold and 53 MeV below $D^0 \bar D^0$ threshold.
A $c\bar{c}$ pair on the color-octet adiabatic surface can be very close to 
the energy shell. The $\psi'$ can therefore  spend a sufficiently long
 time on this surface 
for the $c \bar c$ pair to respond to the repulsive short-distance potential.
This response can lead to a significant suppression of
the $c \bar c$ wavefunction at the origin for the $|c\bar{c}_8 + S \rangle$.

If the initial-state factor ${\cal I}_{J/\psi} (c \bar c_8(^3S_1) + S)$
in (\ref{fact-endpt}) is suppressed for all soft partons $S$,
the suppression can be expressed in the form of 
a relation between the NRQCD matrix elements in (\ref{fact}):
\begin{equation}
{\langle {\cal O}_8(^3S_1) \rangle_{\psi'}
 \over \langle {\cal O}_1(^3S_1) \rangle_{\psi'}}
\; \ll\; 
{\langle {\cal O}_8(^3S_1) \rangle_{J/\psi}
	\over \langle {\cal O}_1(^3S_1) \rangle_{J/\psi}} \;.
\label{ineq}
\end{equation}
This inequality can be tested by 
calculating the matrix elements	using Monte Carlo simulations of 
lattice NRQCD.  Since the $\psi'$ is so close to the $D \bar D$ threshold, 
it would be essential to include dynamical light quarks in the simulations. 

Our proposal leads to a prediction for the flavor-dependence of the 
suppression of the decays of $\psi'$ into vector/pseudoscalar final states.
Bramon, Escribano, and Scadron \cite{B-E-S} 
have analyzed the decays $J/\psi \to VP$ assuming that the decay 
amplitude is the sum of a flavor-connected
amplitude $g$, a flavor-disconnected amplitude $rg$,
and an isospin-violating amplitude $e$.  They allowed for
violations of $SU(3)$ flavor symmetry through parameters $s$ and $x-1$.
Two of these amplitudes are
\begin{eqnarray}
{\cal T}(J/\psi \to \rho \pi) 
&=& (g + e) \; \mbox{\boldmath $\epsilon$}_{J/\psi} \times 
	\mbox{\boldmath $\epsilon$}_\rho \cdot {\bf p}_\pi,
\label{T-rhopi}
\\
{\cal T}(J/\psi \to {K^*}^0 \bar K^0) 
&=& [(1-s)g - (1+x)e] \;
	\mbox{\boldmath $\epsilon$}_{J/\psi} \times 
	\mbox{\boldmath $\epsilon$}_{K^*} \cdot {\bf p}_{\bar K}. 
\end{eqnarray}
The amplitudes for the other $VP$ decay modes 
are given in Ref. \cite{B-E-S}.
The authors give two sets of parameters that fit the existing data,
one with $x=1$ and the other with $x=0.64$.
Both sets have $e$ comparable in magnitude to $rg$ 
and an order of magnitude smaller than $g$. 
If the decay  $J/\psi \to \rho \pi$ 
is dominated by endpoint contributions, we can identify $g$ and $e$ 
with the two terms in (\ref{fact-endpt}).
While $rg$ may also have endpoint contributions from $c \bar c \to g g$,
we assume for simplicity that it is dominated by subasymptotic 
contributions from the $c \bar c_1(^3S_1)$ Fock state. 
The amplitudes for the decays $\psi' \to VP$ 
can be expressed in a similar way in terms of amplitudes
$g'$, $e'$, and $(rg)'$.  Our explanation of the 
$\rho \,$--$\pi$ puzzle implies that $|g'|$ is much smaller than $|g|$,
and that $e'$ and $(rg)'$ differ from $e$ and $rg$ only by the 
factor required by the 15\% rule.
The unknown amplitude $g'$ is constrained by the
BES data on $\psi' \to \rho \pi$ and
$\psi' \to {K^*}^0 \bar K^0$.
The upper bound on $B(\psi' \to \rho \pi)$ gives an upper bound on 
$|g'+e'|^2$, which implies that $g'$ lies in a circle 
in the complex $g'$-plane.  The BES measurement of 
$B(\psi' \to {K^*}^0 \bar K^0)$ gives an allowed range
for $|(1-s)g' - (1+x)e'|^2$, which constrains  $g'$ to an annulus.
The intersection of the interior of the circle with the annulus 
is the allowed region for $g'$.  By varying $g'$ over that region and
taking into account the uncertainties in the parameters of Ref. \cite{B-E-S},
we obtain the predictions for $Q_{VP}$ in Table~1.
Measurements of the $\psi'$ branching fractions consistent 
with these predictions would imply that the suppression of 
the vector/pseudoscalar decays is due to 
the suppression of $g'$.  This
would lend support to our explanation of the $\rho \,$--$\pi$ puzzle.

Our proposal also has implications for the angular distributions of
other two-body decay modes. In general, the angular distribution must 
have the form $1 + \alpha \cos^2 \theta$, with $-1 < \alpha < +1$.
Our solution to the $\rho \,$--$\pi$ puzzle is based on the suppression
of a contribution to $\psi'$ decays that gives the angular distribution 
$1 + \cos^2 \theta$. Thus the parameter $\alpha$ for any two-body decay 
of the $\psi'$ should be less than or equal to $\alpha$ for the 
corresponding $J/\psi$ decay.

A solution to the $\rho \,$--$\pi$ puzzle should also be able to explain 
the pattern  of suppression for various $J^{PC}$ states with the same 
flavors. A preliminary measurement of the axial-vector/pseudoscalar 
decay mode $\psi' \to b_1 \pi$ by the BES collaboration\cite{Olsen} 
gives $Q_{b_1 \pi} = (24 \pm 7) \%$, consistent with no suppression
relative to the $15 \%$ rule. A preliminary measurement of
the vector/tensor decay mode $\psi' \to \rho a_2$  \cite{Olsen} 
gives $Q_{\rho a_2} = (2.9 \pm 1.6) \%$, which, though suppressed
relative to the $15 \%$ rule, is an order of magnitude larger than 
$Q_{\rho \pi}$. This pattern can be explained by also taking into account the 
orbital-angular-momentum selection rule for exclusive amplitudes
in perturbative QCD \cite{Chernyak-Zhitnitsky}.
The decay modes $b_1 \pi$ and $\rho a_2$ both have form factors
that are allowed by the helicity selection rule.
They also both have form factors that violate the helicity selection rule,
but are compatible with an endpoint contribution from 
$c \bar c \to q \bar q$.  However, in the case of $b_1 \pi$, 
the endpoint contribution is further suppressed
by the violation of the orbital-angular-momentum selection rule.
A thorough analysis of $J/\psi$ and $\psi'$
decays into  axial-vector/pseudoscalar and vector/tensor 
final states will be presented elsewhere.

In conclusion, we have proposed a new explanation of the 
$\rho \,$--$\pi$ puzzle. 
We suggest that the decay $J/\psi \to \rho \pi$ is dominated by a 
Fock state in which the $c \bar c$ is in a color-octet $^3S_1$ state
which decays via $c \bar c \to q \bar q$. 
The suppression of this decay mode for the $\psi'$ 
is attributed to a dynamical effect that suppresses the 
$c \bar c$ wavefunction at the origin 
for Fock states that contain a color-octet $c \bar c$ pair.
Our explanation for the $\rho \,$--$\pi$ puzzle
can be tested by studying the flavor dependence of the two-body decay modes 
of the $J/\psi$ and $\psi'$, their angular distributions, and their
dependence on the $J^{PC}$ quantum numbers of the final-state mesons. 

\acknowledgements

This work was supported in part by the U.S.
Department of Energy, Division of High Energy Physics, under
Grant DE-FG02-91-ER40690.
We thank Steve Olsen for discussions of the
data on charmonium decays from the BES collaboration.


%
%
\begin{table}[t]
\label{table-1}
\vspace{5mm}
\begin{center}
\begin{minipage}{2.7in}
\begin{tabular}{|c|c|c|} 
\hline  
$VP$                   & ~~~$x = 1$~~~ & ~~~$x = 0.64$~~~\\ 
\hline 
$ \rho \pi$ 		& $0   - 0.25$	& $0   - 0.25$	\\   
$K^{*0} \bar{K}^0+c.c. $ 	& $1.2 - 2.0$	& $1.2 - 3.0$	\\   
$K^{*+} K^- +c.c. $ 	& $0   - 0.36$	& $0   - 0.52 $	\\   
$\omega \eta $		& $0   - 1.6 $	& $0   - 1.6 $	\\   
$\omega \eta'$ 		& $10  - 51 $	& $12  - 55$	\\   
$\phi \eta$  		& $0.8 - 3.6$	& $0.4 - 3.0$	\\   
$\phi \eta'$  		& $0.7 - 2.5$	& $0.5 - 2.2$	\\   
$ \rho \eta$  		& $14  - 22$	& $14  - 22$	\\   
$ \rho \eta'$  		& $12  - 20$	& $13  - 21$	\\   
$\omega \pi$  		& $11  - 17$	& $11  - 17$	\\   
\hline  
\end{tabular} 
\end{minipage}
\end{center} 
\end{table} 

\noindent
Table 1. Predictions for $Q_{V P}$ in units of $1 \%$ for all the 
	vector/pseudoscalar final states. The values for
	$\rho \pi$ and $K^{*0} \bar{K}^0 +c.c. $  were used as input.
	The columns labelled $x=1$ and $x=0.64$ correspond to the 
	two parameter sets of Ref. \cite{B-E-S}.

\end{document}